\documentclass[a4paper, conference]{IEEEtran}
\IEEEoverridecommandlockouts

\usepackage{cite}
\usepackage{amsmath,amssymb,amsfonts}
\usepackage{graphicx}
\usepackage{textcomp}
\usepackage{xcolor}
\usepackage{mdframed}
\usepackage{enumitem}
\usepackage{comment}
\usepackage{lipsum}
\usepackage{booktabs}
\usepackage{algorithm}
\usepackage[noend]{algpseudocode}
\usepackage{multirow}
\usepackage{tabularx}
\usepackage{url}
\usepackage[most]{tcolorbox}
\usepackage{subcaption}
\usepackage{todonotes}
\usepackage[font=footnotesize,labelfont=bf]{caption}

\algtext*{EndWhile}
\algtext*{EndIf}
\algtext*{EndFor}

\usepackage{bbding}
\usepackage{fontawesome5}
\newcommand{\cmark}{\textcolor{green!70!black}{\faCheckCircle}} 
\newcommand{\xmark}{\textcolor{red!70!black}{\faTimesCircle}} 
\newcommand{\warn}{\textcolor{yellow!60!red}{\faExclamationCircle}}

\usepackage{flushend}
\usepackage[a4paper, total={184mm,239mm}]{geometry}
\def\BibTeX{{\rm B\kern-.05em{\sc i\kern-.025em b}\kern-.08em
    T\kern-.1667em\lower.7ex\hbox{E}\kern-.125emX}}

\newcounter{promptbox}
\renewcommand{\thepromptbox}{\arabic{promptbox}}

\tcbset{
  numberedprompt/.style={
    colback=black!75,
    colframe=black!95,
    coltext=gray!35,
    boxrule=0.6pt,
    arc=5pt,
    left=1pt, right=1pt, top=1pt, bottom=1pt,
    fontupper=\sffamily\footnotesize,
    enhanced,
    before skip=0pt, after skip=0pt,
    before=\par\noindent\ignorespaces,
    after=\unskip\par,
    before upper={\parindent=0pt},
  }
}

\newenvironment{PromptBox}[3][]{%
  \footnotesize
  \refstepcounter{promptbox}%
  \begin{tcolorbox}[numberedprompt,
    title={#2~\thepromptbox: #3}, 
    label={#1}]%
}{\end{tcolorbox}}

\usepackage{listings}
\lstdefinelanguage{yaml}{
  morekeywords={true,false,null,y,n},
  sensitive=false,
  morecomment=[l]{\#},
  morestring=[b]",
  morestring=[b]'
}

\begin{document}

\title{Bench4HLS: End-to-End Evaluation of LLMs in High-Level Synthesis Code Generation}

\author{\IEEEauthorblockN{M Zafir Sadik Khan, Kimia Azar, Hadi Kamali}
\IEEEauthorblockA{\textit{Department of Electrical and Computer Engineering (ECE), University of Central Florida, Orlando, FL 32816, USA} \\
\{mzafirsadik.khan, azar, kamali\}@ucf.edu}
}

\maketitle

\begin{abstract}

In last two years, large language models (LLMs) have shown strong capabilities in code generation, including hardware design at register-transfer level (RTL). While their use in high-level synthesis (HLS) remains comparatively less mature, the ratio of HLS- to RTL-focused studies has shifted from 1:10 to 2:10 in the past six months, indicating growing interest in leveraging LLMs for high-level design entry while relying on downstream synthesis for optimization. This growing trend highlights the need for a comprehensive benchmarking and evaluation framework dedicated to LLM-based HLS. To address this, We present Bench4HLS for evaluating LLM-generated HLS designs. Bench4HLS comprises 170 manually drafted and validated case studies, spanning small kernels to complex accelerators, curated from widely used public repositories. The framework supports fully automated assessment of compilation success, functional correctness via simulation, and synthesis feasibility/optimization. Crucially, Bench4HLS integrates a pluggable API for power, performance, and area (PPA) analysis across various HLS toolchains and architectures, demonstrated here with Xilinx Vitis HLS and validated on Catapult HLS. By providing a structured, extensible, and plug-and-play testbed, Bench4HLS establishes a foundational methodology for benchmarking LLMs in HLS workflows\footnote{The code and resources related to this work are publicly available at: \url{https://github.com/zfsadik/Bench4HLS}}. 
\end{abstract}

\begin{IEEEkeywords}
Large Language Model (LLM), High Level Synthesis (HLS), Design Space Exploration (DSE), Power, Performance, Area (PPA) Efficiency.
\end{IEEEkeywords}

\section{Introduction}

High-level synthesis (HLS) emerged as a solution to the limitations of conventional register-transfer level (RTL) design by enabling automated hardware generation from high-level languages (HLLs) \cite{coussy2010high, lahti2018we}. Its goal is to boost productivity and ease application-specific design space exploration (DSE) without the need for RTL coding knowledge \cite{gajski2012high}. Over the past decade, HLS has matured significantly, supporting rapid design cycles and enabling advanced optimizations in hardware \cite{liu2019accelerating, zhang2021towards, cong2022fpga, cortes2016high}. Nonetheless, achieving optimal power, performance, and area (PPA) remains a challenging, often requiring deep expertise and extensive manual intervention \cite{cong2022fpga, shi2023sechls}.

Now, more than a decade after the emergence of HLS, the rapid rise of large language models (LLMs) and their widespread use in programming tasks have opened new opportunities for hardware design. Recent studies have begun exploring how LLMs can assist in HLS workflows (or even replace portions of them through direct code generation) \cite{swaroopa2024evaluating, xiong2024hlspilot, liao2024llmsgood, xu2024automated, sheikholeslam2024synthai, xu2024optimizing, collini2024c2hlsc, oztas2024agentic, mashnoor2025timelyhls}. These studies largely address isolated parts of the design pipeline, e.g., improving the synthesizability through \textbf{\texttt{pragma}} insertion and modification or code restructuring to be aligned with synthesis tools \cite{xiong2024hlspilot, xu2024optimizing}, transforming (repairing or re-factoring) general-purpose C/C++ code into HLS-compatible kernels through \cite{xu2024automated, collini2024c2hlsc}, and even bypassing traditional HLS flows entirely by directly translating HLLs into RTL via reasoning-based approaches \cite{liao2024llmsgood}.

Despite the growing advancements in utilizing LLMs for HLS, systematic evaluation of such solutions remains limited. To show the gap, on the other side, RTL-level evaluation is already supported by established efforts such as VerilogEval \cite{Liu2023verilogeval}, RTLLM \cite{rtllm}, and more recently CVDP benchmark \cite{pinckney2025comprehensive}. While some argue that direct LLM-to-RTL generation may reduce the role of HLS, this overlooks the fact that \textbf{\textit{HLS tools are deeply optimized for FPGA-based acceleration and provide essential abstraction layers for DSE}} \cite{schafer2019high}. Thus, LLM-assisted HLS remains a critical research direction, and having a fully automated, systematic evaluation benchmark is a must. A recent benchmark, HLS-Eval \cite{hlseval2025}, has taken first steps in this space, but its scope is restricted: While it emphasizes syntactic and synthesizability checks, it overlooks PPA analysis and relies only on small-scale kernels. Hence, there remains a gap in robust frameworks that holistically evaluate LLM-generated HLS code across the full synthesis pipeline, including quality-of-results (QoR) considerations essential for practical deployment \cite{5737854}.

To address this gap, we propose Bench4HLS, a comprehensive evaluation framework designed to benchmark LLMs in HLS workflows. While Table \ref{tab:top_comparison} reflects the main advantages of Bench4HLS vs. the prior art, our contributions are as follows:

\noindent \underline{\textit{(i) Diverse Benchmark Suite:}} Bench4HLS includes 170 curated and verified HLS design tasks spanning small kernels to accelerator-scale applications, collected from widely used open-source repositories and standardized with golden testbenches for reliable reproducibility. 

\noindent \underline{\textit{(ii) Pluggable PPA evaluation:}} Bench4HLS integrates a modular API for PPA analysis across multiple HLS toolchains, shown (as the main testcase) here with Xilinx Vitis HLS and successfully tested on Catapult HLS.

\noindent \underline{\textit{(iii) Design Space Exploration (DSE):}} Bench4HLS explores synthesis-level (tool-in-the-loop) exploration within its evaluation loop, exposing Pareto frontiers under fixed design budgets.

\noindent \underline{\textit{(iv) Rigorous correctness validation:}} Bench4HLS enforces rigorous functional verification at multiple stages, pre-synthesis C simulation, HLS co-simulation, and post-implementation netlist validation, ensuring that only timing-clean, correctness-preserving designs are scored. 

By combining dataset breadth, multi-objective evaluation, and automated optimization, Bench4HLS provides the first holistic and practically meaningful benchmark for assessing LLM-driven and automated HLS design methodologies\footnote{Benchmarking models and scripts will be released upon paper acceptance.}.

\section{Background and Related Work}

\begin{figure*}[t]
\centering
\includegraphics[width=\linewidth]{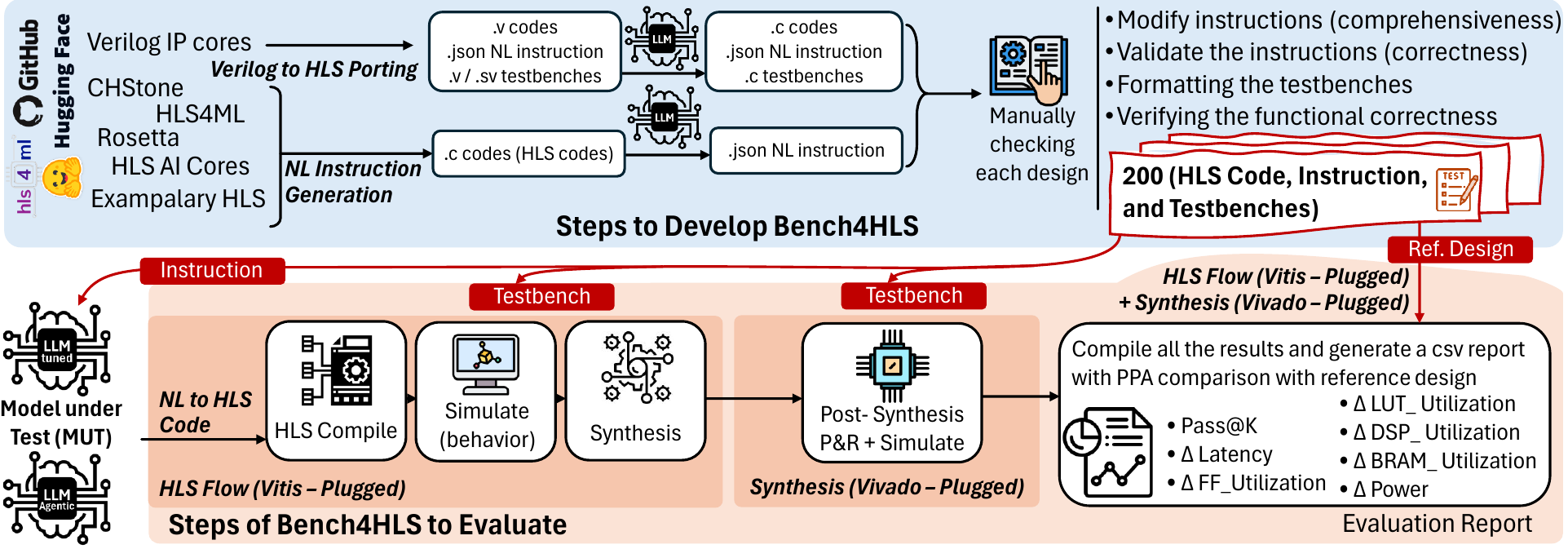}
\vspace{-10pt}
\caption{Overview of Bench4HLS: The whole workflow of the evaluation method}
\vspace{-15pt}
\label{fig:bench4hls_framework}
\end{figure*}

\subsection{LLMs for Direct RTL Code Generation}

RTL is the most widely studied hardware target for LLM-based code generation \cite{surveyllmeda}, from fine-tuning \cite{liu2024rtlcoder, akyash2025rtl} to agentic \cite{zhao2025mage, ping2025hdlcore} and decoding \cite{akyash2025decortl} approaches. In terms of benchmarking and evaluation, several benchmarks established de-facto baselines. VerilogEval evaluates LLMs on HDLBits-style problems and, more recently, specification-to-RTL tasks with improved harnesses; it remains a common point of comparison \cite{10.1145/3718088}. RTLLM extends beyond syntax/functional checks to include a “design quality” goal and provides an open repository of tasks and testbenches \cite{rtllm}.

\subsection{LLMs for HLS Code Generation and Synthesis}

Unlike RTL, there is a relatively limited prior studies focusing on the use of LLM for HLS. HLS reliance on implicit compiler optimizations, hierarchical transformations, and synthesis constraints, challenges LLMs that must generate syntactically correct code while reasoning about structural, timing, and resource constraints \cite{prakriya2025lift}. These ``LLMs for HLS" efforts can be grouped into four key categories:

\begin{table}[b]
\centering
\setlength{\tabcolsep}{2.5pt}
\caption{Comparison of LLM-based HLS Evaluation Frameworks.}
\label{tab:top_comparison}
\begin{tabular}{@{} l *{4}c @{}}
\toprule
\textbf{Dimension}  & \textbf{Gai \textit{et al} \cite{gai2025exploring}} & \textbf{HLS-Eval \cite{hlseval2025}} & \textbf{Bench4HLS (prop.)} \\
\cmidrule(r){1-1} \cmidrule(r){2-2} \cmidrule(r){3-3} \cmidrule(r){4-4}
\textbf{\# of Benchmarks} & 52 & 94 & 170 \\
\cmidrule(r){1-1} \cmidrule(r){2-2} \cmidrule(r){3-3} \cmidrule(r){4-4}
\textbf{Benchmarks} & Small to & Small to & Small to \\ 
\textbf{Size} & Medium-sized  & Medium-sized & Large-sized \\
\textbf{(LoC$^*$/Benchmark}\textbf{)} & (56) & (43) & (88) \\
\cmidrule(r){1-1} \cmidrule(r){2-2} \cmidrule(r){3-3} \cmidrule(r){4-4}
\multirow{2}{*}{\textbf{Simulation}} & \multirow{2}{*}{\warn~Pre-HLS} & \cmark~Pre- and & \cmark~Pre-, Post-HLS \\ 
 &  & Post-HLS & and Post-P\&R \\
\cmidrule(r){1-1} \cmidrule(r){2-2} \cmidrule(r){3-3} \cmidrule(r){4-4}
\textbf{Tool (Synthesis)} &  \warn~Partial & \warn~Fixed Tool & \cmark~Pluggable  \\
& Support & on Vitis & (Vitis/Catapult)\\
\cmidrule(r){1-1} \cmidrule(r){2-2} \cmidrule(r){3-3} \cmidrule(r){4-4}
\textbf{PPA Analysis} & \xmark~No & \xmark~No & \cmark~Yes \\
\cmidrule(r){1-1} \cmidrule(r){2-2} \cmidrule(r){3-3} \cmidrule(r){4-4}
\textbf{DSE} & \xmark~No & \xmark~No & \cmark~YAML-based \\
\cmidrule(r){1-1} \cmidrule(r){2-2} \cmidrule(r){3-3} \cmidrule(r){4-4}
\textbf{Code Generation} & \xmark~No & \cmark~Yes & \warn~Not Optimized$^+$ \\
\cmidrule(r){1-1} \cmidrule(r){2-2} \cmidrule(r){3-3} \cmidrule(r){4-4}
\textbf{Open-Source} & \xmark~No & \cmark~Yes & \cmark~Yes \\
\cmidrule(r){1-1} \cmidrule(r){2-2} \cmidrule(r){3-3} \cmidrule(r){4-4}
\textbf{Parallel Test} & \xmark~No & \cmark~Yes & \cmark~Yes \\
\bottomrule
\multicolumn{4}{l}{$^*$: LoC/Benchmark: Average \# of Code Lines in each Benchmark.} \\
\multicolumn{4}{l}{$^+$: Can generate but not optimized as the main aim of Bench4HLS.}
\end{tabular}
\end{table}

\noindent \textit{\underline{(1) Synthesizability and PPA Optimization}}. Initial studies show that LLMs can generate functionally correct HLS code, but without hardware-aware optimizations, performance and area lag behind expert designs \cite{liao2024llmsgood}. Feedback-driven approaches such as HLSPilot improve results by incorporating profiling and synthesis constraints during generation \cite{xiong2024hlspilot}.

\noindent \textit{\underline{(2) Code Refactoring and Repair}}. LLMs have been applied to transform general-purpose C/C++ into HLS-ready kernels, addressing unsupported constructs and improving synthesizability. While effective for basic transformations, challenges remain for complex optimizations such as hierarchical architectures and memory dataflow \cite{collini2024c2hlsc, xu2024automated}.

\noindent \textit{\underline{(3) Structured Reasoning and Agents}}. Frameworks like SynthAI \cite{sheikholeslam2024synthai}, LIFT \cite{prakriya2025lift} and agentic optimization methods extend LLMs with reasoning, graph analysis, and tool interaction. These systems automate pragma insertion, modularize tasks, and iteratively refine implementations, improving synthesis success rate and design quality \cite{collini2025reasoningmodelsreasonhardware}.

\noindent \textit{\underline{(4) Dataset and Model Design}}. Specialized datasets and model architectures have been introduced to enhance HLS support. For example, SAGE-HLS leverages AST-guided training on synthetic HLS-C corpora, achieving high synthesis accuracy but still limited functional correctness \cite{khan2025sage}.

\subsection{Benchmarking of LLM Generated HLS Codes}

Benchmarking for LLM-based HLS evaluation is still at an early stage, with only a few targeted efforts: (1) Gai et al. \cite{gai2025exploring} introduce one of the first benchmarks coupling natural-language prompts with HLS designs. Their framework focuses on syntactic and semantic correctness checks, focusing on reliable automated HLS code generation from text specifications; (2) HLS-Eval \cite{hlseval2025} provides an extensible Python-based infrastructure that integrates with commercial HLS tools. It offers a curated suite of designs with testbenches and natural-language prompts, enabling user-configurable evaluation of LLM-generated HLS code; (3) HLSTester \cite{xu2025hlstester} augments evaluation by detecting functional mismatches between C/C++ source and synthesized hardware. By combining LLM reasoning with adaptive fuzzing and guided input generation, it improves test coverage and error detection efficiency.

Existing efforts remain narrow in scope, where they primarily assess syntactic or functional correctness, while overlooking hardware-critical metrics such as PPA, as well as its DSE. This gap motivates the development of Bench4HLS, which extends beyond correctness to incorporate QoR metrics and multi-tool evaluation, offering a more rigorous and practically meaningful benchmark for LLM-based HLS. A detailed comparison with prior frameworks is shown in Table \ref{tab:top_comparison}.

\section{Proposed Framework: Bench4HLS}

To systematically evaluate LLM-based HLS code generation, Bench4HLS consists of four primary stages: (i) dataset construction, (ii) code generation via LLMs, (iii) functional verification and synthesis using Vitis HLS, and (iv) performance characterization and comparative analysis which is demonstrated in Figure~\ref{fig:bench4hls_framework}. Throughout these stages, we form an end-to-end pipeline capable of assessing not only syntactic and semantic correctness, but also hardware-level metrics that are essential for real-world deployment.

\subsection{Dataset: Instructions, Designs, and Testbenches}

The first stage involves constructing a curated dataset composed of 170 designs at various scales, each represented as a triplet: a natural language instruction, a synthesizable HLS (C/C++) implementation, and a corresponding HLL (C/C++) testbench. The triplets were constructed as follows:

\noindent \textbf{\textit{\underline{(i) Reference Design Collection:}}} We curated a comprehensive, multiscale dataset of heterogeneous test cases. It spans from small-scale Verilog datasets (e.g., VerilogEval \cite{Liu2023verilogeval}), Vitis-HLS-Introductory-Examples \cite{XilinxVitisHLS}, and HLS codes from textbook references \cite{fingeroff2010hls,kastner2018parallel}, to mid- and large-scale accelerators and arithmetic circuits from established repositories such as CHStone \cite{chstone}, HLS4ML \cite{fahim2021hls4ml}, and Rosetta \cite{zhou2018rosetta}. This diversity yields an average of 88 lines of code (LoC) per test case (see Table \ref{tab:top_comparison}), establishing the largest and most representative dataset of its kind to date. 

\noindent \textbf{\textit{\underline{(ii) Formatting and Instruction Preparation:}}} For the selected Verilog test cases, we ported each design into HLS-C++ using the GPT-5 model, followed by manual verification to ensure correctness (see Prompt~\ref{box:v2hls} for an example). In total, 170 HLS codes were produced, from which GPT-5 subsequently generated natural language instructions for each design. To enable comprehensive evaluation, corresponding testbenches were also constructed and manually validated, ensuring the reliability and completeness of the dataset.

\begin{PromptBox}[box:v2hls]{Prompt}{Verilog to HLS (C/C++) Porting.}
```\\
You are an expert in HLS-C++ and Verilog code generation. I will give you a prompt for a hardware design written for verilog, the reference code in verilog and it's testbench.\\

Your task is to generate the reference design and the testbench for HLS-C++. Your generated testbench must reflect the same logic of the reference code.\\

All your generated codes will be run in HLS tools. While writing the testbenches consider all possible scenario. Your testbench has to be self checking. Write the codes without the 'RefModule'. \\

The testbench will detect errors of design file but it must not show compilation and simulation error. It should also always return 0. Use all necessary and applicable pragmas for better optimization in your design code.\\

The three items for verilog are below:\\
\\\{Prompt\}
\\\{Reference Code\}
\\\{Testbench\} \\
'''
\end{PromptBox}

\noindent \textbf{\textit{\underline{(iii) Manual Dataset Verification:}}} Each instruction was validated across multiple LLM prompting (self-refinement \cite{madaan2023self}) to confirm its robustness and faithfulness to the design intent. The testbenches were constructed to mirror functional coverage and contain embedded logic derived from the corresponding design. The correctness of each triplet was verified through manual inspection and simulation. While manual, all 170 entries of our curated dataset is tailored for both functional and PPA evaluation of instruction-to-HLS generation tasks.

\subsection{LLM-Based HLS Code Generation}

Although efficient code generation is not the primary objective of Bench4HLS, the framework is designed with a flexible, pluggable API that enables the integration of LLMs for HLS code generation across its 200-task dataset. At this stage, the supported natural language instructions serve as task prompts, which are passed through the API to the integrated LLMs for producing the corresponding HLS code implementations. Since no open-source LLMs are currently dedicated to HLS, we employ widely used open- and closed-source coding models, e.g., GPT-5, QwenCoder, and Llama, as case studies. 

Using this pluggable model demonstrates how Bench4HLS can be applied in realistic scenarios, where a model-under-test (MUT) is systematically evaluated for functionality, synthesizability, as well as PPA efficiency through the framework’s evaluation pipeline.

\noindent \textbf{\textit{\underline{(i) Plugging Model (API):}}} For testing phase, we integrated and evaluated several models, including proprietary and open-source models such as GPT-5, QwenCoder, and Llama. The Python API (vLLM and simonw/llm Python libraries) is fully  standardized for instruction-based inputs and code-oriented responses, abstracting away vendor-specific differences (especially w.r.t. HLS tools differences in syntax).

\noindent \textbf{\textit{\underline{(ii) Prompting and Parsing:}}} Each model was queried using the instruction generated by the prompt described in Prompt ~\ref{box:instruction2hls}, which directed the model to produce high-level, system-design explanations of the HLS C++ module. Outputs were post-processed to remove commentary, headers, or test scaffolding, ensuring that the retained content focused solely on the module’s functionality, dataflow, pipelining, memory behavior, and control logic. Custom scripts further standardized formatting and terminology for consistency across results.

\begin{PromptBox}[box:instruction2hls]{Prompt}{Instruction (Spec.) Generation for HLS Test Cases.}
```\\
You are an expert in writing and explaining HLS-C++ hardware deisgn. Describe the functionality of the HLS design from a high-level, system-design perspective. \\

Focus on what the module computes, how data flows through it, and how control logic operates across different stages of execution. Explain its functional behavior considering pipelining, memory access, and synthesis constraints. \\

Avoid unnecessary low-level details like variable names or line-by-line mappings. Instead, communicate the purpose, the algorithmic steps, the I/O behavior, and any state or timing dependencies using precise, practical language that would guide someone implementing or verifying the design. \\

Include timing models, pipeline stages, and basic handshaking logic if relevant, but keep the explanation focused and hardware-aware.\\ \\
\{Reference HLS-C++ Code\}\\
'''
\end{PromptBox}

\noindent \textbf{\textit{\underline{(ii) Dataset Organization:}}} The extracted designs were stored using a systematic naming scheme, enabling seamless batch processing in downstream compilation and evaluation steps. This stage produces a structured set of LLM-generated HLS codes that serve as the main subject for subsequent functional and performance evaluation.

\subsection{Automated Verification and Synthesis via Vitis HLS}

The Bench4HLS framework provides a fully automated flow for evaluating high-level synthesis. By design, the framework is pluggable, allowing different HLS and implementation toolchains to be seamlessly integrated. In the current instantiation, the front end is connected to Vitis HLS for compilation, simulation, and synthesis, while the back end is linked to Vivado for post-RTL placement and routing, enabling detailed PPA analysis. This integration delivers a holistic view of design quality, combining cycle-accurate semantic validation with synthesis-proven correctness. Importantly, the framework is architected for batch-mode evaluation, supporting parallel execution across multiple designs. This capability is particularly critical as HLS flows can be computationally demanding for complex benchmarks. The evaluation proceeds through three principal stages: \noindent \textit{\underline{(1) Compilation:}} Designs is compiled to detect syntactical errors. Failing designs are recorded as compilation failures. \noindent \textit{\underline{(2) Pre-Synthesis Simulation:}} compiled designs are simulated using the corresponding testbenches. This verifies the functional behavior of the generated code against expected outputs. \noindent \textit{\underline{(3) Synthesis and Post-Synthesis Simulation:}} Designs that pass simulation are synthesized to obtain RTL-level implementations. This is followed by RTL synthesis for resource utilization metrics (e.g., LUTs, FFs, BRAMs), and then post-synthesis simulation for cycle-accurate functional verification. 

All steps are executed through an automated script pipeline comprising batch files and TCL scripts. This ensures reproducibility and scalability of the benchmarking process across hundreds of designs and models. This stage acts as a filter that identifies which LLM-based generated codes are not only syntactically valid, but also functionally correct and suitable for hardware implementation at acceptable PPA.

\begin{algorithm}[t]
\scriptsize
\caption{Automated Evaluation w/ DSE in Bench4HLS.}
\label{alg:automation_dse}
\begin{algorithmic}[1]

\Require HLS design set $\mathcal{D}$, testbench set $\mathcal{T}$, Pass@K, exploration policy $\pi$, part $\mathsf{FPGA}$.
\Ensure Structured report $\mathcal{R}$ with per-solution status and PPA metrics.

\State Initialize $\mathcal{R} \gets \emptyset$
\State Partition $\mathcal{D}$ into $K$ batches: $\{\mathcal{D}_1,\dots,\mathcal{D}_K\}$ \Comment{Top-$K$ / Pass@K batching}
\For{$r \gets 1$ \textbf{to} $K$}
  \For{$d_i \in \mathcal{D}_r$}
    \State Retrieve testbench $t_i \in \mathcal{T}$ and YAML DSE spec $Y_i$
    \State Expand $Y_i$ into solution set $\Omega_i \gets \textsc{Expand}(Y_i,\pi)$
    \State Initialize per-design results $\mathcal{R}_i \gets \emptyset$
    \For{each solution $s \in \Omega_i$}
      \State Open Plugged HLS project with $d_i$ and apply directives from $s$ 
      \State \textsc{csim\_design} $\rightarrow$ compilation status $\sigma^{\text{com}}_{i,s}$
      \State Append $(d_i,s,\sigma^{\text{com}}_{i,s})$ to $\mathcal{R}_i$
      \If{$\sigma^{\text{com}}_{i,s}=\texttt{PASS}$}
        \State Simulate $d_i$ with $t_i$ in Plugged HLS $\rightarrow$ status $\sigma^{\text{sim}}_{i,s}$
        \State Append $(d_i,s,\sigma^{\text{sim}}_{i,s})$ to $\mathcal{R}_i$
        \If{$\sigma^{\text{sim}}_{i,s}=\texttt{PASS}$}
          \State \textsc{csynth\_design} $\rightarrow$ status $\sigma^{\text{syn}}_{i,s}$, latency $L_{i,s}$, area vector $A_{i,s}$
          \State Append $(d_i,s,\sigma^{\text{syn}}_{i,s},L_{i,s},A_{i,s})$ to $\mathcal{R}_i$
          \State Export RTL $\rightarrow$ Synth\_Tool; set strategy from $s$ 
          \State Parse $\mathsf{FPGA}$ timing/power $\rightarrow$ $\text{WNS}_{i,s}$, $F_{\max\,i,s}$, $P_{i,s}$
          \State Append $(d_i,s,\text{WNS}_{i,s},F_{\max\,i,s},P_{i,s})$ to $\mathcal{R}_i$
        \EndIf
      \EndIf
    \EndFor
    \State Append $(d_i,\mathcal{R}_i,\Pi_i)$ to $\mathcal{R}$
  \EndFor
\EndFor

\State Compute Pass@K across stages using $\{\sigma^{\text{com}}_{i,s},\sigma^{\text{sim}}_{i,s},\sigma^{\text{syn}}_{i,s}\}$ aggregated per design 
\State Append Pass@K summary rows and per-design bests to $\mathcal{R}$
\State \Return $\mathcal{R}$

\end{algorithmic}
\end{algorithm}

\subsection{DSE-based Characterization and Comparative Analysis}

Bench4HLS evaluates LLM-generated HLS designs not only in terms of functional correctness but also through detailed PPA metrics\footnote{All PPA-related results are compared with its corresponding reference implementation (provided by Bench4HLS) using normalized PPA metrics.}. Primarily, the framework integrates DSE to assess how plugged compiler directives and implementation strategies affect these outcomes. Some of the main metrics evaluated in Bench4HLS are (i) latency (in nanoseconds), (ii) hardware resource utilization (LUTs, FFs, DSPs, BRAMs), (iii) power analysis of synthesized RTL. Additionally, Bench4HLS computes Pass@K metrics across compilation, simulation, and synthesis stages.

\begin{table*}[t]
\footnotesize
\centering
\setlength\tabcolsep{11pt}
\caption{Pass@K evaluation of Compilation, Simulation and Synthesis Status for selected Large Language Models}
\label{tab:pass@k_table}
\begin{tabular}{@{} l *{21}c @{}}
\toprule
\multirow{2}{*}{Evaluated Model} & \multicolumn{3}{c}{Compilation Status} & \multicolumn{3}{c}{Simulation Status} & \multicolumn{3}{c}{Synthesis Status} \\
\cmidrule(r){2-4} \cmidrule(r){5-7} \cmidrule(r){8-10}
& Pass@1 & Pass@5 & Pass@10 & Pass@1 & Pass@5 & Pass@10 & Pass@1 & Pass@5 & Pass@10 \\
\cmidrule(r){1-1} \cmidrule(r){2-4} \cmidrule(r){5-7} \cmidrule(r){8-10}
Qwen 2.5 Coder 14B & 47.06\% & 58.82\% & 65.88\% & 18.82\% & 25.88\% & 30.59\% & 17.65\% & 24.71\% & 28.82\%  \\
\cmidrule(r){1-1} \cmidrule(r){2-4} \cmidrule(r){5-7} \cmidrule(r){8-10}
Qwen 2.5 Coder 32B & 82.35\% & 90.59\% & 92.35\% & 28.24\% & 36.06\% & 39.41\% & 24.12\% & 34.71\% & 38.24\% \\
\cmidrule(r){1-1} \cmidrule(r){2-4} \cmidrule(r){5-7} \cmidrule(r){8-10}
Llama 3.3 70B & 77.65\% & 95.88\% & 97.65\% & 45.29\% & 58.24\% & 64.12\% & 44.12\% & 57.65\% & 63.53\% \\
\cmidrule(r){1-1} \cmidrule(r){2-4} \cmidrule(r){5-7} \cmidrule(r){8-10}
GPT-5 & 85.29\% & 97.65\% & 97.65\% & 52.35\% & 65.88\% & 72.35\% & 50.00\% & 65.88\% & 71.76\% \\
\bottomrule
\end{tabular}
\vspace{-10pt}
\end{table*}

Additionally, each design can be associated with a declarative YAML specification that enumerates the design space to be further explored. Bench4HLS sweeps through this space by instantiating multiple solutions (serial or in parallel), applying directives in plugged HLS (e.g., Vitis HLS), collecting synthesis reports, and comparing the reports for DSE. The overview of this automated process is outlined in Algorithm~\ref{alg:automation_dse} Various directives and pragmas will be included w.r.t. the specification of design, e.g., pipelining, tunable intervals, dataflow, unrolling, etc. As shown in Listing \ref{lst:yaml_dse}, a detailed YAML specification has been used in Bench4HLS for this process:

\begin{lstlisting}[language=yaml,caption={Example YAML configuration for DSE in Bench4HLS},label={lst:yaml_dse}]
clock_period_ns: [3.3, 5.0]
enable_pipeline: [true, false]
pipeline_ii: [1, 2]
enable_dataflow: [true, false]
unroll_factor: [1, 2, 4, 8]
array_partition_factor: [1, 2, 4]
allocation_limit_add: [0, 1, 2]   # 0 = disabled
dsp_full_reg: [true, false]
vivado_strategy: [Default, Performance_Explore, Area_Explore]
\end{lstlisting}

By exposing the DSE space in a structured, machine-readable YAML format, Bench4HLS enables reproducible and extensible DSE. This design makes Bench4HLS not only a benchmarking platform, but also a practical framework for systematic DSE studies of LLM-generated HLS designs.

\begin{PromptBox}[box:hlsgen]{Prompt}{Instruction used for Prompting the Model for Test.}
```\\
Your role is to act as an expert in HLS C++ code development. You must thoroughly explore each question through a systematic and deliberate thinking process—engaging in cycles of analysis, summarization, exploration, reassessment, reflection, backtracing, and iteration to develop well-considered solutions. \\

Based on the provided instructions, you are expected to generate precise, optimized, and accurate synthesizable HLS C++ code that meets the following requirements.\\

        1. Output only synthesizable C++ code (no extra text or explanation).\\
        2. Include necessary headers (e.g. ap\_int.h).\\
        3. The top-level function signature must be exactly as given in the instruction.\\
        4. Insert suitable HLS pragmas.\\
        5. Self-contained: do not rely on external libraries or files beyond standard headers.\\
        
    \#\#\# Instruction:\{\}\\
    \#\#\# Input:\{\}\\
    \#\#\# Response:\{\}\\
'''
\end{PromptBox}

\section{Experimental Setup}

To demonstrate the comprehensiveness of Bench4HLS, we evaluated the HLS code generation of four LLMs: Qwen2.5-Coder (14B and 32B), LlaMA 3.3 70B, and GPT-5. The Qwen and LlaMA models were deployed locally with 4-bit quantization for efficient inference, while GPT-5 was accessed via OpenAI’s API. For each of the 170 benchmark tasks, we used a structured prompt mentioned in Prompt~\ref{box:hlsgen} that explicitly instructed the model to produce synthesizable, pragma-annotated HLS C++ code (combined with YAML for DSE). While pluggable, for this paper and at post-generation and synthesis stage, all designs were evaluated using Vitis HLS 2024.1 for compilation, simulation, and synthesis, and Vivado 2024.1 for power estimation. All designs targeted the Xilinx Artix-7 xc7a200tffv1156-1 FPGA to ensure consistent comparison. The entire workflow of Bench4HLS is automated using custom pluggable scripts with TCL drivers, enabling scalable evaluation across multiple models and Pass@K configurations.

\section{Experimental Results}

We evaluated Bench4HLS using four state-of-the-art LLMs: \textbf{Qwen2.5-Coder 14B}, \textbf{Qwen2.5-Coder 32B}, \textbf{LlaMA 3.3 70B}, and \textbf{GPT-5}. The goal is to assess the ability of Bench4HLS for evaluating the capacity of given models for translating natural language descriptions into functionally correct and synthesizable HLS designs. Bench4HLS provides a fully automated pipeline for design generation, simulation, synthesis, and PPA analysis, enabling reproducible and comprehensive benchmarking. The evaluation flow is organized as a four-stage pipeline: (i) \textbf{Syntactic validity}: compilation success via \textsc{csim\_design}; (ii) \textbf{Semantic correctness}: functional validation using testbench-driven simulation; (iii) \textbf{Hardware feasibility}: synthesizability through RTL generation in Vivado; and (iv) \textbf{DSE}: automated exploration of HLS directives and Vivado strategies using YAML specifications (see Listing~\ref{lst:yaml_dse}) for optimum PPA. For each stage, Bench4HLS computes Pass@K metrics ($K \in \{1,5,10\}$), measuring the likelihood that at least one of the top-$K$ generated candidates passes. Table~\ref{tab:pass@k_table} summarizes the results across compilation, simulation, and synthesis stages. Bench4HLS allows us to integrate any new MUT to be evaluated based on this four-stage pipeline. 
 
Bench4HLS provides model-wise comparison as shown in Table ~\ref{tab:pass@k_table}. For this examplary case, \textbf{GPT-5} achieves the strongest performance, with 97.65\% compilation success at Pass@10, and 72.35\% and 71.76\% success in simulation and synthesis respectively. \textbf{LlaMA 3.3 70B} follows closely, while the \textbf{Qwen2.5} series shows the impact of scaling: moving from 14B to 32B yields significant improvements in syntactic and semantic fidelity. These results highlight the sensitivity of HLS-oriented code generation to model size and capacity.

\begin{figure*}
\centering
\includegraphics[width=\linewidth]{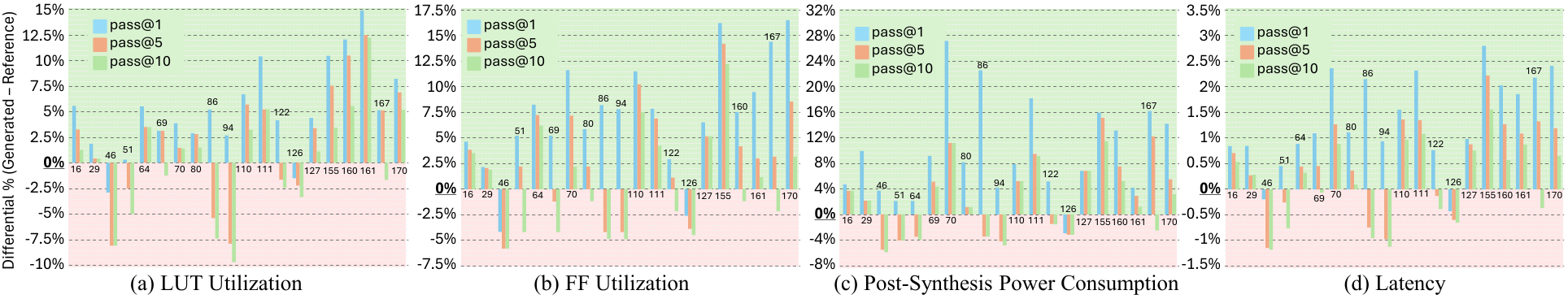}
\caption{PPA Analysis of LLM-generated (GPT) HLS Codes by Bench4HLS. Bench4HLS reports percentage differences between the generated designs by MUT and the reference designs provided by Bench4HLS across four metrics (LUT utilization, FF utilization, post-synthesis power, and latency -- for pass@1, pass@5, and pass@10 generations), with values representing (generated – reference)\%. Positive means higher resource usage or latency vs. the reference.}
\vspace{-15pt}
\label{fig:gpt_diff_ppa}
\end{figure*}

\begin{figure}
\centering
\includegraphics[width=\linewidth]{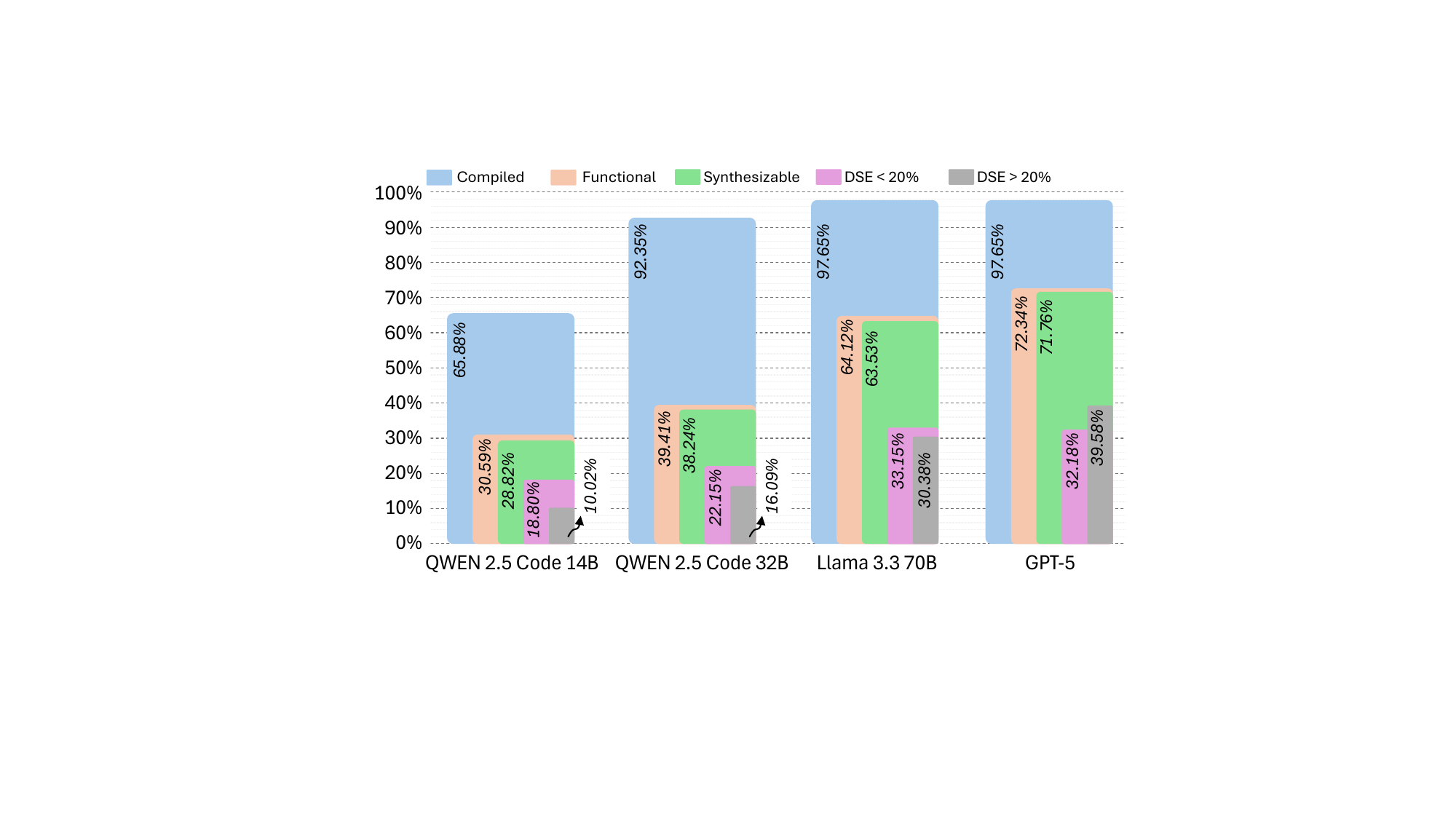}
\caption{HLS Code Generation Evaluated by Bench4HLS for Pass@10 (DSE refers to improvement of at least one of PPA metrics less/more than 20\%).}
\label{fig:avg_percentile}
\end{figure}

To further reflect the capability of Bench4HLS, automated YAML-based HLS DSE has been evaluated here. As shown in Figure \ref{fig:avg_percentile}, for four given model, Bench4HLS can simulate and synthesize benchmarks with configurable parameter (as exemplified in Listing \ref{lst:yaml_dse}). Figure \ref{fig:avg_percentile} reflects the improved results while DSE is applied (in this case, when at least one of PPA metrics are improved less/more than 20\%). As shown, for GPT-5, DSE improves at least one PPA metric by $>20\%$ in $\sim40\%$ of benchmarks, compared to only $\sim10\%$ for Qwen2.5-14B. This demonstrates the importance of directive-level search for realizing the full potential of LLM-generated designs. This automated DSE allows us to exhaustively test the MUT effectiveness across various configurations. 

Additionally, beyond HLS synthesis, Bench4HLS integrates post-RTL implementation in given (pluggable)  synthesis tool (here is Vivado) using a configurable target FPGA (here is \textit{Xilinx Artix-7 xc7a200tffv1156-1}). Four primary metrics are extracted relative to reference designs: latency (\% of improvement vs. reference), FF utilization (\% of improvement vs. reference), LUT utilization (\% of improvement vs. reference), and estimated power consumption (\% of improvement vs. reference). Figure \ref{fig:gpt_diff_ppa} demonstrates these metrics for the designs generated by GPT-5. A key strength of Bench4HLS lies in its curated set of reference designs, which serve as a high-quality baseline for comparison. As shown in Figure~\ref{fig:gpt_diff_ppa}, the reference implementations consistently outperform even the strongest model (GPT-5) in terms of PPA efficiency. For smaller benchmarks (problems \#1--\#100), LLM-generated solutions achieve metrics close to the reference baseline, but for larger and more complex benchmarks (\#150+), most models struggle to produce synthesizable designs. In selected designs, the resulting PPAs are significantly worse than the reference, underscoring the efficiency of the curated designs. This trend is even more highlighted for smaller models such as Qwen2.5-14B/32B and LlaMA 3.3 70B, which fail more frequently and diverge further from the baseline. These results highlight that the reference designs in Bench4HLS are not merely functional baselines, but \textbf{well-optimized, efficiency-oriented implementations} that establish a meaningful upper bound against which automated approaches can be evaluated. 

\begin{figure}[t]
\centering
\includegraphics[width=\linewidth]{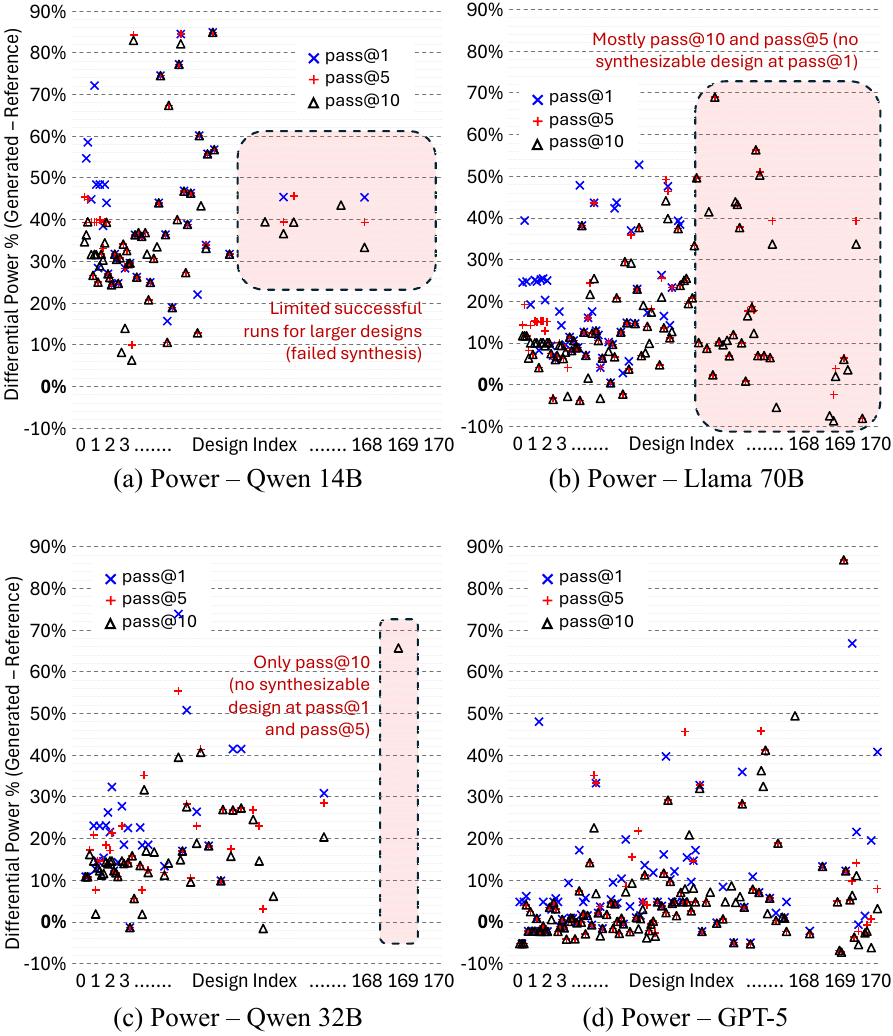}
\caption{Power Consumption Benchmarking in MUTs Tested by Bench4HLS.}
\label{fig:power_models}
\vspace{-5pt}
\end{figure}

Figure \ref{fig:power_models} reveals this scalability issue better, where power distributions across all benchmarks have been illustrated. As shown, smaller models (e.g., Qwen2.5-14B) fail to produce synthesizable designs for many larger problems (sparse results beyond design \#120), while GPT-5 produces more valid RTL for a broader range of cases. The breadth of successful synthesis reflects Bench4HLS’s automated pipeline for consistent PPA extraction and synthesis validation. These experiments demonstrate that Bench4HLS provides an effective and extensible framework for evaluating LLM-generated HLS designs. By combining Pass@K correctness evaluation with YAML-driven DSE and post-RTL PPA analysis, the framework enables holistic comparison across models.

\section{Conclusion}

This paper proposes Bench4HLS, a comprehensive benchmarking and evaluation framework dedicated to assessing LLMs for HLS code generation. By combining a curated set of 170 optimized reference designs with an automated four-stage evaluation flow using pluggable industry-standard tools for compilation, simulation, synthesis, and YAML-based DSE, Bench4HLS enables systematic variation of HLS directives and backend strategies for rigorous measurement of correctness, feasibility, and PPA efficiency. To assess the capabilities of Bench4HLS, four advanced LLMs have been evaluated, i.e., GPT-5, Llama 3.3 70B, and Qwen 2.5 14B/32B. By offering an extensible, plug-and-play testbed, Bench4HLS establishes a foundation for the systematic study of LLMs in HLS design.

\bibliographystyle{IEEEtran}
\bibliography{reference}

\end{document}